\documentstyle[psfig]{mn}

\newif\ifAMStwofonts

\ifoldfss
  \ifCUPmtlplainloaded \else
    \NewTextAlphabet{textbfit} {cmbxti10} {}
    \NewTextAlphabet{textbfss} {cmssbx10} {}
    \NewMathAlphabet{mathbfit} {cmbxti10} {} 
    \NewMathAlphabet{mathbfss} {cmssbx10} {} 
  \fi
  \ifAMStwofonts
    \ifCUPmtlplainloaded \else
      \NewSymbolFont{upmath} {eurm10}
      \NewSymbolFont{AMSa} {msam10}
      \NewMathSymbol{\upi}     {0}{upmath}{19}
      \NewMathSymbol{\umu}     {0}{upmath}{16}
      \NewMathSymbol{\upartial}{0}{upmath}{40}
      \NewMathSymbol{\leqslant}{3}{AMSa}{36}
      \NewMathSymbol{\geqslant}{3}{AMSa}{3E}

       \let\le=\leqslant
       \let\ge=\geqslant
    \fi
  \fi
\fi 

\ifnfssone
  \newmathalphabet{\mathit}
  \addtoversion{normal}{\mathit}{cmr}{m}{it}
  \addtoversion{bold}{\mathit}{cmr}{bx}{it}
  \newmathalphabet{\mathbfit} 
  \addtoversion{normal}{\mathbfit}{cmr}{bx}{it}
  \addtoversion{bold}{\mathbfit}{cmr}{bx}{it}
  \newmathalphabet{\mathbfss} 
  \addtoversion{normal}{\mathbfss}{cmss}{bx}{n}
  \addtoversion{bold}{\mathbfss}{cmss}{bx}{n}
  \ifAMStwofonts
    \ifCUPmtlplainloaded \else
      \UseAMStwoboldmath
      \makeatletter
      \new@mathgroup\upmath@group
      \define@mathgroup\mv@normal\upmath@group{eur}{m}{n}
      \define@mathgroup\mv@bold\upmath@group{eur}{b}{n}
      \edef\UPM{\hexnumber\upmath@group}
      \new@mathgroup\amsa@group
      \define@mathgroup\mv@normal\amsa@group{msa}{m}{n}
      \define@mathgroup\mv@bold\amsa@group{msa}{m}{n}
      \edef\AMSa{\hexnumber\amsa@group}
      \makeatother
      \mathchardef\upi="0\UPM19
      \mathchardef\umu="0\UPM16
      \mathchardef\upartial="0\UPM40
      \mathchardef\leqslant="3\AMSa36
      \mathchardef\geqslant="3\AMSa3E

       \let\le=\leqslant
       \let\ge=\geqslant
    \fi
  \fi
\fi 

\ifnfsstwo
  \DeclareMathAlphabet{\mathbfit}{OT1}{cmr}{bx}{it}
  \SetMathAlphabet\mathbfit{bold}{OT1}{cmr}{bx}{it}
  \DeclareMathAlphabet{\mathbfss}{OT1}{cmss}{bx}{n}
  \SetMathAlphabet\mathbfss{bold}{OT1}{cmss}{bx}{n}
  \ifAMStwofonts
    \ifCUPmtlplainloaded \else
      \DeclareSymbolFont{UPM}{U}{eur}{m}{n}
      \SetSymbolFont{UPM}{bold}{U}{eur}{b}{n}
      \DeclareSymbolFont{AMSa}{U}{msa}{m}{n}
      \DeclareMathSymbol{\upi}{0}{UPM}{"19}
      \DeclareMathSymbol{\umu}{0}{UPM}{"16}
      \DeclareMathSymbol{\upartial}{0}{UPM}{"40}
      \DeclareMathSymbol{\leqslant}{3}{AMSa}{"36}
      \DeclareMathSymbol{\geqslant}{3}{AMSa}{"3E}

       \let\le=\leqslant
       \let\ge=\geqslant
    \fi
  \fi
\fi 

\ifCUPmtlplainloaded \else
  \ifAMStwofonts \else 
    \def\upi{\pi}
    \def\umu{\mu}
    \def\upartial{\partial}
  \fi
\fi

\title[Formation of intragroup HI rings]{Massive HI clouds with no optical 
  counterparts as high-density regions of intragroup HI rings and arcs} 
\author[K. Bekki,  B. S. Koribalski, S. D. Ryder,  W.  J. Couch]
       {K. Bekki${}^1$, B. S. Koribalski${}^2$, S. D. Ryder${}^3$, 
	and W. J. Couch${}^1$\\
        ${}^1$School of Physics, University of New South Wales, 
	      Sydney 2052, NSW, Australia \\
        ${}^2$Australia Telescope National Facility, CSIRO, P.O. Box 76, 
	      Epping NSW 1710, Australia\\
        ${}^3$Anglo-Australian Observatory, P.O. Box 296, Epping NSW 1710, 
	      Australia}
\date{Accepted,
      Received,
      in original form June 2004}

\pagerange{\pageref{firstpage}--\pageref{lastpage}}
\pubyear{2004}

\begin{document}

\maketitle

\label{firstpage}

\begin{abstract}

We present a new scenario in which massive intragroup HI clouds are the 
high-density parts of large HI rings/arcs formed by dynamical interaction 
between galaxy groups and gas-rich, low surface brightness (LSB) galaxies
with extended gas disks. 
Our hydrodynamical simulations demonstrate that the group tidal field is 
very efficient at stripping the outer HI gas of the disk if the gaseous 
disk of the LSB galaxy extends $2 - 5$ times further than the stellar disk.
We find that  a massive, extended `leading stream' orbiting 
the group's center can form out of the stripped outer HI envelope, while 
the severely shrunk LSB galaxy, whose stellar disk remains unaffected, 
continues on its path. The result is a relatively isolated, 
massive HI cloud with a ring- or arc-like shape, a very inhomogeneous 
density distribution ($N_{\rm HI} \sim 1.0 \times 10^{17} - 1.1 \times 10^{20}$
atoms\,cm$^{-2}$), and, initially, no stellar content. 
Only the high density peaks of the simulated intragroup HI ring/arc can 
be detected in many current HI observations. These will appear as relatively
isolated `HI islands' near the group center. 
We also find that star formation can occur within the ring/arc,
if the total gas mass within the intragroup ring/arc is very large 
($\sim$ 4 $\times$ $10^9$ ${\rm M}_{\odot}$).
We discuss these results in terms of 
existing observations of intragroup gas (e.g., the Leo Ring and HIPASS 
J0731--69) and intergalactic HII regions.
 
\end{abstract}


\section{Introduction}


\begin{table*}
\centering
\begin{minipage}{185mm}
\caption{Model parameters and results on intragroup ring formation}
\begin{tabular}{ccccccccll}
No. 
& $M_{\rm s}$  ($\times$ $10^{8}$  ${\rm M_{\odot}}$)  
& $M_{\rm dm}/M_{\rm s}$ 
& $M_{\rm g}/M_{\rm s}$ 
& $R_{\rm s}$ (kpc)
& $R_{\rm g}/R_{\rm s}$ 
& $X_{\rm g}$ (kpc)
& $V_{\rm g}$  
& model description 
& ring characteristics \\
M1 & 7.8 &  20  &  4  &  6.4 & 5 & 200 & 0.75$V_{c}$ & fiducial & ring/arc \\
M2 & 7.8 &  20  & 10  &  6.4 & 5 & 200 & 0.75$V_{c}$ & more massive & rings with star formation \\
M3 & 7.8 &  20  &  4  &  4.0 & 5 & 200 & 0.75$V_{c}$ & HSB disk & no remarkable rings  \\
M4 & 7.8 &  20  &  4  &  6.4 & 2 & 200 & 0.75$V_{c}$ & compact HI disk  & no remarkable rings\\
M5 & 7.8 &  20  &  4  &  6.4 & 5 & 200 & 0.50$V_{c}$ & smaller pericenter & more massive rings \\
\end{tabular}
\end{minipage}
\end{table*}

Massive intragroup HI gas clouds with apparently no optical counterparts are 
known to exist in several galaxy systems (for overviews see, e.g., Hibbard 
et al. 2001, Koribalski 2004). While many of these clouds are clearly part of
tidal streams/bridges formed by and still connected to interacting or merging 
galaxies, like for example the Magellanic Clouds (Putman et al. 2003), the  
M\,81 group (Yun et al. 1994), the NGC~3256 group (English et al. 2004), and 
IC~2554 (Koribalski et al. 2003), only a few are found at relatively large 
distances from their apparent progenitors. The most spectacular example in 
the latter category is the massive, intergalactic HI gas cloud, HIPASS 
J0731--69 ($v_{\rm hel} \approx 1480$ km\,s$^{-1}$), in the NGC~2442 galaxy 
group (Ryder et al. 2001). This extended cloud, which has an HI mass of 
$\sim10^9 M_{\odot}$ and lies at a projected separation of 180 kpc from the 
gas-rich, asymmetric spiral galaxy NGC~2442\footnote{We adopt a distance of
$D$ = 15.5 Mpc for the NGC~2442 galaxy group (see Ryder et al. 2001).}, was 
discovered in the HIPASS Bright Galaxy Catalog (Koribalski et al. 2004) and 
is the only definite extragalactic HI cloud in this catalog. 

Surrounding the elliptical galaxy NGC~1490, Oosterloo et al. (2004) found 
several intragroup HI clouds with a total HI mass of $\sim10^{10}$ $M_{\odot}$ 
distributed along an arc 500 kpc long. Large clumpy HI rings (diameter 
$\sim200$ kpc) have also been detected around the E/S0 galaxies M\,105 
('The Leo Ring', Schneider et al. 1989), 
NGC~5291 (Malphrus et al. 1997; Duc \& Mirabel  1998) and 
NGC~1533 (Ryan-Weber et al. 2003).
These are in contrast to the much smaller, smooth HI rings (diameter $\sim25$
kpc) with regular velocity fields around NGC~2292/3 (Barnes 1999; Barnes \&
Webster 2001) and IC\,2006 (Franx et al. 1994). 

The purpose of this paper is to propose
that there is a physical/evolutionary link between the massive isolated
HI gas without apparent optical counterparts (e.g., HIPASS J0731-69)
and the intragroup HI ring (e.g., the Leo ring).
Using numerical simulations,
we first demonstrate that intragroup HI rings or arcs can be formed from 
the outer gaseous envelope of a galaxy interacting with a group potential.
Then we show that only the highest-density regions of the intragroup 
rings/arcs are generally observable in HI emission. 
In the following discussions,
the theoretical 5$\sigma$ HI column density of 
HIPASS (for  a velocity width of 100 km\,s$^{-1}$)
is $4 \times 10^{18}$ atoms\,cm$^{-2}$ for extended objects filling
the beam (see Koribalski et al. 2004). 

\begin{figure}
\psfig{file=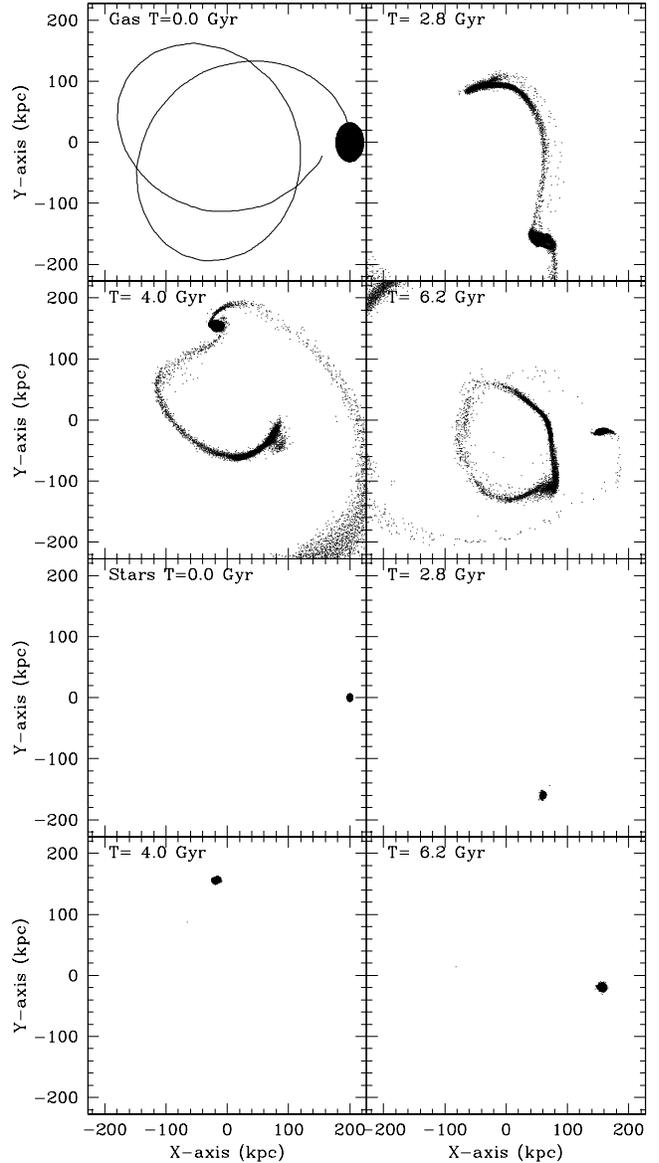,width=8.5cm}
\caption{ 
 Morphological evolution of a low surface brightness (LSB) galaxy with an {\em 
 extended HI disk} orbiting the center of the group in the fiducial model 
 (M1 in Table~1), projected onto the $x$-$y$ plane. The gaseous component 
 is shown at the top, the stellar component at the bottom. The center of the 
 group is at ($x,y$) = (0,0) for all panels. The orbital evolution of the LSB 
 galaxy is given by a solid line in the upper left panel. $T$ represents the 
 time that has elapsed since the start of the simulation. Note that a giant 
 intragroup HI ring of size $\sim$200 kpc is formed at $T$ = 6.2 Gyr and  
 contains no stars. 
The size,  the location with respect to the center of the group,
 the total mass,  and the rotational kinematics of the ring
at $T$ = 6.2 Gyr 
are similar to those of the Leo ring (Schneider et al. 1989). 
}
\label{Figure. 1}
\end{figure}

\begin{figure*}
\psfig{file=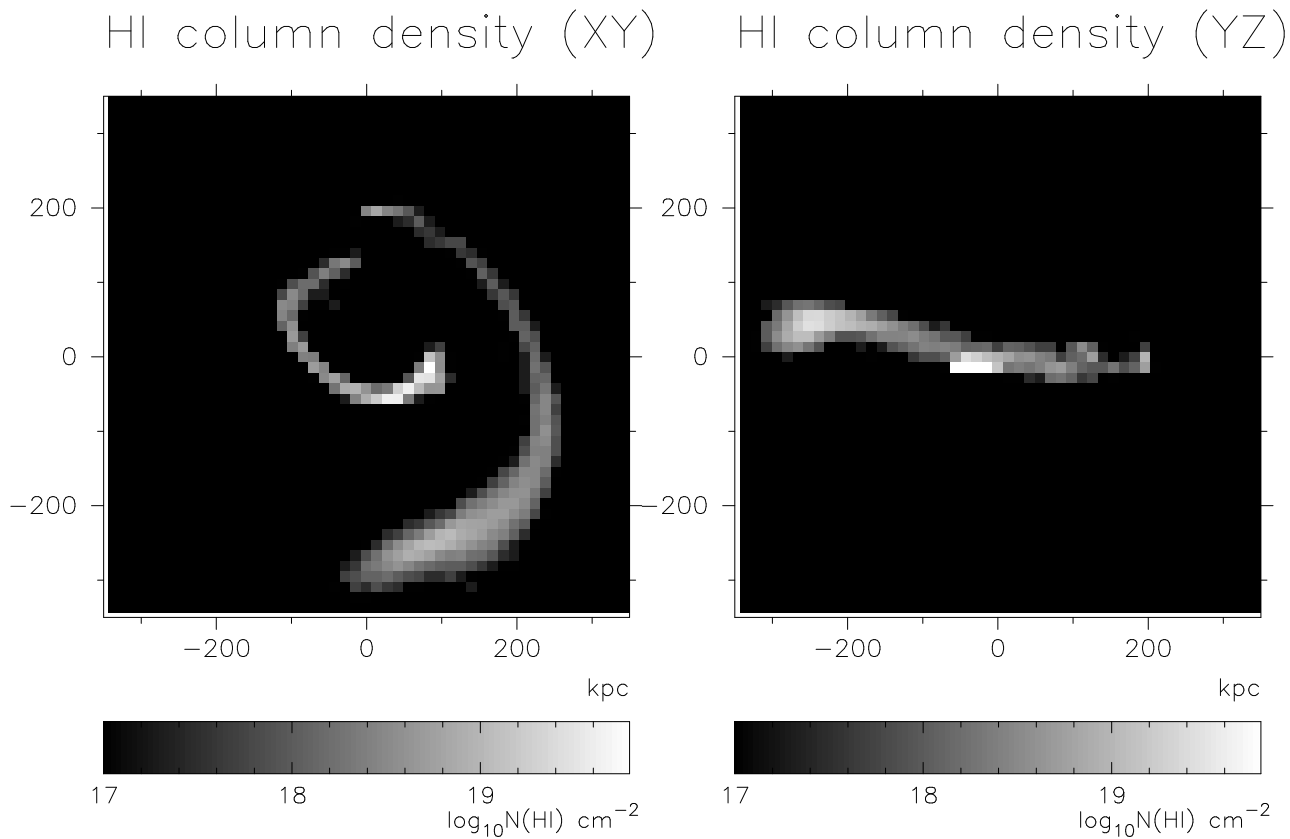,width=17.0cm}
\caption{ 
 HI column density distribution ($N_{\rm HI}$) of the simulated intragroup 
 ring/arc projected onto the $x$-$y$ plane (left) and the $y$-$z$ plane (right)
 in the fiducial model at $T$ = 4.0 Gyr. Here only `stripped gas particles' 
 that are outside the initial HI gas disk of the LSB galaxy are used to 
 estimate the column density. The cell size is 28 kpc. 
 Cells with $N_{\rm HI} \le 
 10^{17}$ atoms\,cm$^{-2}$ (including those without gas particles, i.e., 
 zero column density) are shown in black. 
}
\label{Figure. 2}
\end{figure*}

\begin{figure*}
\psfig{file=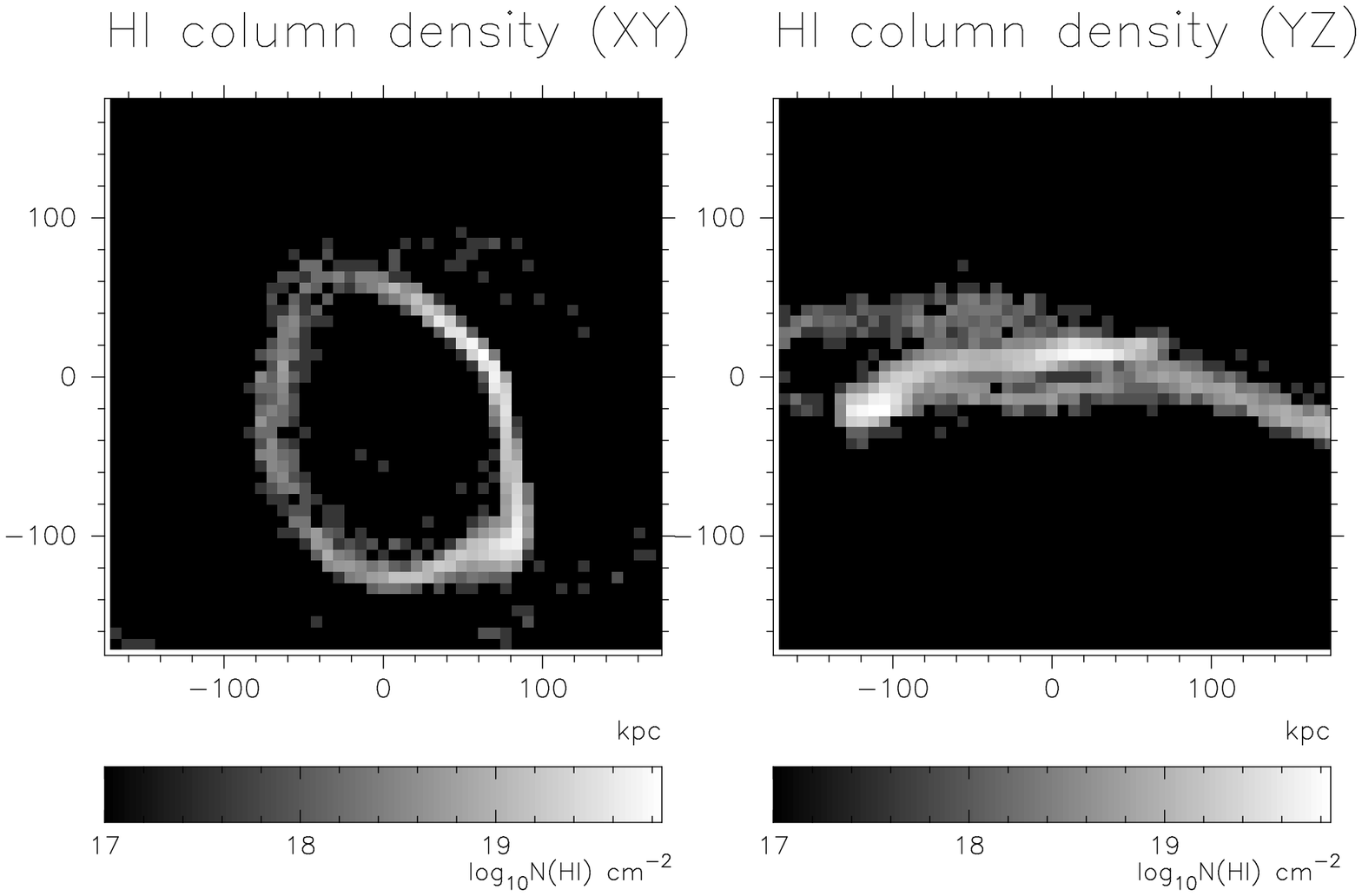,width=17.0cm}
\caption{ 
The same as Fig. 2 but for $T$ = 6.2 Gyr.
}
\label{Figure. 3}
\end{figure*}

\section{The model}

We consider a bulgeless late-type  galaxy (here a LSB)
with an extended HI gas disk orbiting 
the center of the group and thereby numerically investigate the evolution of 
the outer HI disk.
The galaxy orbit is determined solely by the fixed external gravitational 
field of the group with the scale-free logarithmic potential of $\Phi \equiv 
{V_{\rm c}}^2 {\rm log}(r)$, where $r$ denotes the distance of the galaxy 
from the group center. 
The adopted smooth potential with no substructures of the group
might be an oversimplified assumption.
Since we intend to discuss later the isolated HI gas
discovered in the NGC~2442 galaxy group,
we choose a reasonable and realistic  value of $V_{c}$ for this group.
By assuming  that the group center is 
nearly coincident with the center of the central elliptical
galaxy NGC 2434 in this group,
we choose 300 km s$^{-1}$ (Rix et al. 1997) as a reasonable $V_{c}$.
The initial orbital plane of the galaxy is the same 
as the $x$-$z$ plane and the stellar and gaseous disk are initially inclined 
${\theta}_{\rm d}$ degrees with respect to the $x$-$y$ plane. The initial 
position and velocity of the galaxy are therefore ($X_{\rm g}$, 0, 0,) and
(0, $V_{\rm g}$, 0), respectively, where $X_{\rm g}$ and $V_{\rm g}$ are 
free parameters which determine the orbit of the galaxy. For most of models, 
${\theta}_{\rm d} = 45^{\circ}$, $X_{\rm g}$ = 200 kpc, and $V_{\rm g}$ = 
0.75$V_{c}$ (see Table~1).

A late-type disk with the total stellar 
mass $M_{\rm s}$ is assumed to be embedded 
in a massive dark matter halo with a universal `NFW' profile (Navarro et al. 
1996) and the mass $M_{\rm dm}$. The disk consists of an exponential {\em 
stellar disk} of size $R_{\rm s}$, radial scale length 0.2 $R_{\rm s}$, and 
mass $M_{\rm s}$ as well as a uniform (or exponential) {\em gas disk} of 
size $R_{\rm g}$ and mass $M_{\rm g}$. 
$M_{\rm s}$, $M_{\rm dm}/M_{\rm s}$, 
$M_{\rm g}/M_{\rm s}$, $R_{\rm s}$, and $R_{\rm g}/R_{\rm s}$ are important 
free parameters in the present study. Star formation in gas disk is modeled 
according to the Schmidt law (Schmidt 1959) with the exponent of 1.5.
The coefficient of the law is chosen such that the mean star formation
rate of the Galactic disk model with gas mass fraction
of 0.1 is $\sim$ 1 ${\rm M_{\odot}}$ yr$^{-1}$
for $\sim$ 1 Gyr dynamical evolution
of the disk (e.g., Bekki et al. 2002). 
These values in the star formation model are consistent with
observations (e.g., Kennicutt 1998). 

We mainly present the results of the `fiducial' model (M1 in Table~1)
with $M_{\rm s} = 7.8 \times 10^{8} {\rm M_{\odot}}$ (corresponding to 
$M_{\rm B} = -16$ mag), $M_{\rm dm}/M_{\rm s}$ = 20, $M_{\rm g}/M_{\rm s}$ 
= 4, $R_{\rm s}$ = 4 kpc (i.e., a $B$-band central surface brightness of 
24.5 mag\,arcsec$^{-2}$), and $R_{\rm g}/R_{\rm s}$ = 5, i.e. a  
LSB galaxy with an HI disk five times larger than the stellar 
disk.

The HI diameters of gas-rich galaxies are generally observed to be larger 
than their optical disks (Broeils \& van Woerden 1994). A small fraction 
of low luminosity galaxies have HI gas envelopes extending out to 4--7 
$R_{\rm s}$ (e.g., Hunter 1997) with HI mass to light ratios up to $\sim 
20 {\rm M_{\odot}\,L_{\odot}^{-1}}$ (Warren et al. 2004).
Therefore, the above parameter set can be regarded as reasonable.
 Parameter values of other representative models are shown in Table~1.
All models include the hydrodynamical evolution of the
isothermal gas; they are investigated using TREESPH codes described in 
Bekki (1997). The resolution of each simulation is $\sim$ 1.6 $\times$
$10^5 {\rm M_{\odot}}$ for mass and $\sim$ 200 pc for scale.
Individual time step is allocated for each particle with the maximum
time step width of 1.4 $\times$ $10^6$ yr.

\section{Results}
Fig. 1 describes how gaseous intragroup rings/arcs are formed from the 
galaxy-group interaction in the fiducial model. As the LSB galaxy passes 
by the pericenter of its orbit for the first time, the group's strong 
tidal field efficiently strips the outer HI disk of the LSB galaxy ($T$ 
= 2.8 Gyr). The stripped HI disk or `leading stream' can orbit the group 
center and consequently forms an arc-like structure with a size of 
$\sim$200 kpc with a very inhomogeneous density distribution ($T$ = 4.0 Gyr).
The total mass of the HI arc (i.e., stripped HI) is $1.4 \times 10^{9} 
M_{\odot}$ (43\% of the initial HI mass) with only 11\% of the initial 
gas converted into new stars mostly within the optical disk ($T$ = 4.0 Gyr).
The projected surface gas density ${\Sigma}_{\rm g}$ 
(or column density $N_{\rm HI}$) ranges from $1.0 \times 10^{17}$ 
atoms cm$^{-2}$
to $1.1 \times 10^{20}$ atoms cm$^{-2}$ along the inner arc.
The HI disk size of the LSB galaxy
is dramatically reduced (from $\sim$5 $R_{\rm s}$ to $\sim$2 $R_{\rm s}$)
because of the tidal stripping of the outer part of the HI disk.

The stellar disk of the LSB galaxy, on the other hand, is less susceptible 
to the tidal field of the group than the gas so that stellar tidal tails are 
not formed in the galaxy-group interaction. Finally, an HI gas ring without 
stars is formed around the center of the group ($T$ = 6.2 Gyr). It should be 
stressed here that the LSB galaxy, from which the HI ring originated, is 
located well outside the ring so that there appears to be no physical 
connection between them. Fig.~2 shows the HI column density distribution of 
the ring at $T$ = 4.0 Gyr.
Since HI column density along the ring/arc is very inhomogeneous,
only the highest density regions such as  the tip of the ring/arc
can be detected in currently ongoing observations for both face-on 
and edge-on views. 
For example,
the number fraction of cells with ${\Sigma}_{\rm g}$ 
$\ge$  4 $\times$ $10^{18}$ atoms cm$^{-2}$ 
corresponding roughly to the HIPASS detection limit
is only 23 \%  among  all cells with  ${\Sigma}_{\rm g}$ 
$\ge$ $10^{17}$ atoms cm$^{-2}$.
However, Fig. 2 also indicates that if the detection limit 
in future observations is as good as $\sim$ $10^{17}$ atoms\,cm$^{-2}$, 
evidence for the ring/arc structures should be revealed.
Fig. 3 also demonstrates that the intragroup ring
has a very inhomogeneous density distribution,
which is consistent with observations of some intragroup
HI rings such as the Leo ring (e.g., Schneider et al. 1989).

Physical properties of the simulated intragroup rings/arcs 
in other 4 representative models (M2$-$M5) are described as follows.
Firstly, in the model M2 where the LSB galaxy initially has a larger
amount of HI gas in the disk ($\sim 8 \times 10^{9} {\rm M_{\odot}}$), 
stars form in parts of the HI rings/arcs 
(See Fig. 4).
This  is due essentially
to the increased HI gas density in the ring as compared to the model M1. 
HI clouds containing young and new stars
can be identified as intergalactic HII regions.
Secondly, intergalactic HI rings/arcs {\it with no optical counterparts}
(i.e., without stellar streams) can neither be  seen  in
the model M3 with HSB disk nor in M4 with more compact HI disk.  
This is 
because (1) tidal stripping of the outer HI gas is much less efficient
in this HSB-group interaction
and (2) most of the gas is consumed by star formation
due to initial high density of gas within the disks 
(i.e., little gas can be stripped).
Thirdly, the larger amount of gas is stripped to form a more massive
ring in the model M5 with smaller pericenter of the LSB. 
This implies that total masses of intragroup HI rings/arcs formed from
galaxy-group interaction depend strongly on the orbits
of galaxies with respect to the group center for a given galactic mass.
Furthermore, the HI disk sizes of galaxies with smaller pericenter are
more significantly reduced during galaxy-group interaction.

\begin{figure}
\psfig{file=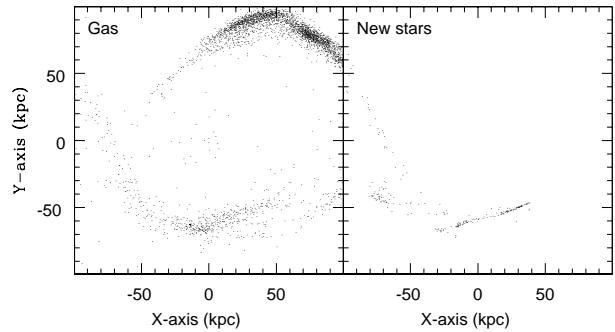,width=8.cm}
\caption{ 
Mass distributions of gas (left) and new stars formed from gas (right)
along the intragroup ring
projected onto the $x$-$y$ plane at $T$ = 4.5 Gyr
for the model M2 (see Table~1) with the 
initial gas mass 2.5 times larger than that of the fiducial model. 
Note that gas density appears to be lower in the gaseous regions 
where young stars are located (i.e., in the lower HI arc).
}
\label{Figure. 4}
\end{figure}

\section{Discussions and Conclusions}

We have demonstrated that (1) gaseous intragroup rings/arcs can be formed 
by dynamical interaction between LSB galaxies {\em with extended HI disks}
and the group potential and (2)  due to their inhomogeneous density 
distribution only the high-density peaks within these rings/arcs can 
currently be observed as apparently isolated clumpy HI clouds. This `tip 
of the iceberg' scenario can not only explain why the observed massive 
intragroup HI clouds (e.g., Ryder et al. 2001) do not have any detectable 
optical counterparts but also provide a clear explanation for the origin 
of HI rings such as the Leo ring (Schneider et al. 1989). This scenario 
predicts that there would be less massive intragroup HI rings/arcs that 
have not been identified as `rings/arcs' in observational studies for HI 
gas of groups of galaxies because of their low HI gas density. 
{\it Some} high velocity clouds (HVCs) observed in the Local Group 
and other galaxy groups can be high-density parts of intragroup rings/arcs.

The present numerical models predict the existence of isolated 
intergalactic star-forming regions located within the high density parts 
of intragroup (intergalactic) HI rings/arcs formed from galaxy--group 
(galaxy--galaxy) interaction.
These star-forming regions, which do not have an old stellar population, are 
in striking contrast to `tidal dwarf galaxies'.
Both theory and observations suggested that
tidal dwarfs are formed in the tidal 
tails of interacting/merging galaxies
thus have old and new stars but no dark matter (e.g., 
Duc \& Mirabel 1998; Duc et al. 2000).
Recently, Ryan-Weber et al. (2004) found a number of very small isolated HII 
regions at projected distances up to 30 kpc from their nearest galaxy (e.g.,
NGC~1533), but located well within the tidal HI features. Oosterloo et al. 
(2004) also found HII regions associated with the HI clouds near NGC~1490.
In the Leo Ring (Schneider et al. 1989) no H$\alpha$ has so far been detected
(Donahue et al. 1995).

Our simulations furthermore
suggest that  the size of extended outer HI disks in galaxies is related to
their environment and interaction history.
Although both sizes and morphological properties of outer HI gas disks are 
observed to vary significantly  between gas-rich galaxies (e.g., Broeils \& 
van Woerden 1994; Broeils \& Rhee 1997), the origin of this diversity 
remains elusive (Broeils \& Rhee 1997). The present numerical results have 
shown that the size of the HI disk of a galaxy interacting with its host 
group can be significantly reduced and the reduction rate depends on the 
galactic orbit. These results suggest that sizes of the outer HI gas disks 
in galaxies can be ``fossil records'' of past dynamical interaction
with its host group. 

The present study suggests that future high sensitivity HI observations
(e.g., the upgraded Arecibo telescope with HI detection limit of
at least $10^{17}$ atoms\,cm$^{-2}$) 
will reveal faint intragroup rings/arcs in which the observed massive HI gas 
clouds are embedded.
Extensive structural and kinematical studies on the intragroup rings/arcs
connected to the isolated clouds will enable us to give strong 
constraints on the orbital evolution of galaxies from which HI gas
was stripped due to group-galaxy interaction.
Thus future observational studies on intragroup HI rings/arcs,
combined with numerical simulations with variously different sets of
orbital parameters of galaxies in groups,
will provide valuable information on the roles of galaxy-group interaction in
galaxy evolution in groups of galaxies.

\section{Acknowledgment}
We are  grateful to the referee for valuable comments,
which contribute to improve the present paper.
KB  and WJC acknowledge the financial support of the Australian Research 
Council throughout the course of this work.
All of the simulations described here were performed on 
NEC SX5/2 systems at AC3 
(Australian Centre for Advanced Computing and Communications).


\end{document}